\documentclass{article}
\usepackage{ijcai26}

\usepackage{times}
\usepackage{soul}
\usepackage{url}
\usepackage[hidelinks]{hyperref}
\usepackage[utf8]{inputenc}
\usepackage[small]{caption}
\usepackage{graphicx}
\usepackage{amsmath}
\usepackage{amsthm}
\usepackage{booktabs}
\usepackage{algorithm}
\usepackage{algorithmic}
\usepackage[switch]{lineno}

\usepackage{multirow}

\title{DaV-Gen: End-to-End Generative Retrieval via Draft-and-Verify}

\author{
Meng Zhao\and
Chunmei Liu\and
Qinyong Wang
\affiliations
HUJING Digital Media \& Entertainment Group\\
\emails
\{chuming.zm, meijiang.lcm, wangqinyong.wqy\}@alibaba-inc.com
}

\begin{document}

\maketitle

\begin{abstract}
Mainstream industrial information retrieval systems (e.g., search and recommendation) are usually built upon Multi-Stage Cascade Architectures (MCAs), which balance effectiveness and efficiency through a coarse-to-fine ``retrieval-ranking'' pipeline.
However, the optimization objectives across different stages are substantially inconsistent, propagating or even amplifying the early-stage errors that ultimately degrade the quality of final results.
While emerging end-to-end generative models offer a potential solution by unifying the pipeline, their online serving performance is severely hindered by the auto-regressive process inherited from the standard decoder-only structure.
To bridge this gap, we introduce \textbf{DaV-Gen}, a novel unified solution designed to fundamentally refactor the paradigm for both search and recommendation via a ``Draft-and-Verify'' mechanism.
Inspired by the process used by speculative decoding, our framework redesigns the generation task into two synergistic operations within a single model.
During training, the model is concurrently optimized for both candidate drafting and fine-grained verification.
This is achieved by a composite loss function that jointly trains the model on two distinct but related objectives: 1) a contrastive loss that structures the embedding space for efficient drafting, and 2) a fusion loss that combines generative likelihood with vector similarity to produce a superior verification score.
This integrated training strategy equips the model with dual capabilities.
At inference time, it first performs highly efficient vector-based drafting to generate a candidate set, and then verifies these candidates using the more powerful fused scoring function, thereby achieving both the speed of sparse drafting and the precision of advanced generative models within a unified, end-to-end architecture.
\end{abstract}

\section{Introduction}

Most large-scale information retrieval systems, including Search Engines and Recommender Systems, rely on a Multi-Stage Cascade Architecture (MCA)~\cite{covington2016deep,zhou2018deep} to balance efficacy and efficiency.
However, this distinct ``retrieve-and-rank'' pipeline suffers from \textit{objective inconsistency}, where misalignment between stage-specific goals leads to \textit{error propagation}---early misses by the retriever are irreversible, fundamentally limiting final result quality~\cite{zhang2025killing,hron2021component}.
To unify this pipeline, recent end-to-end models~\cite{deng2025onerec} represent items as discrete ``Semantic IDs''~\cite{rajput2023recommender}, transforming retrieval into a sequence-to-sequence task.
Despite their promise, standard Generative Information Retrieval (GenIR) models are severely hindered by their auto-regressive nature, which introduces prohibitive inference latency and lacks deterministic control over recommendation list length.

To bridge the gap between the efficiency of cascades and the unified expressiveness of generative models, we introduce \textbf{DaV-Gen}, a novel framework that refactors the paradigm via a \textbf{Draft-and-Verify} mechanism.
\begin{figure*}[t]
    \centering
    \includegraphics[width=\textwidth]{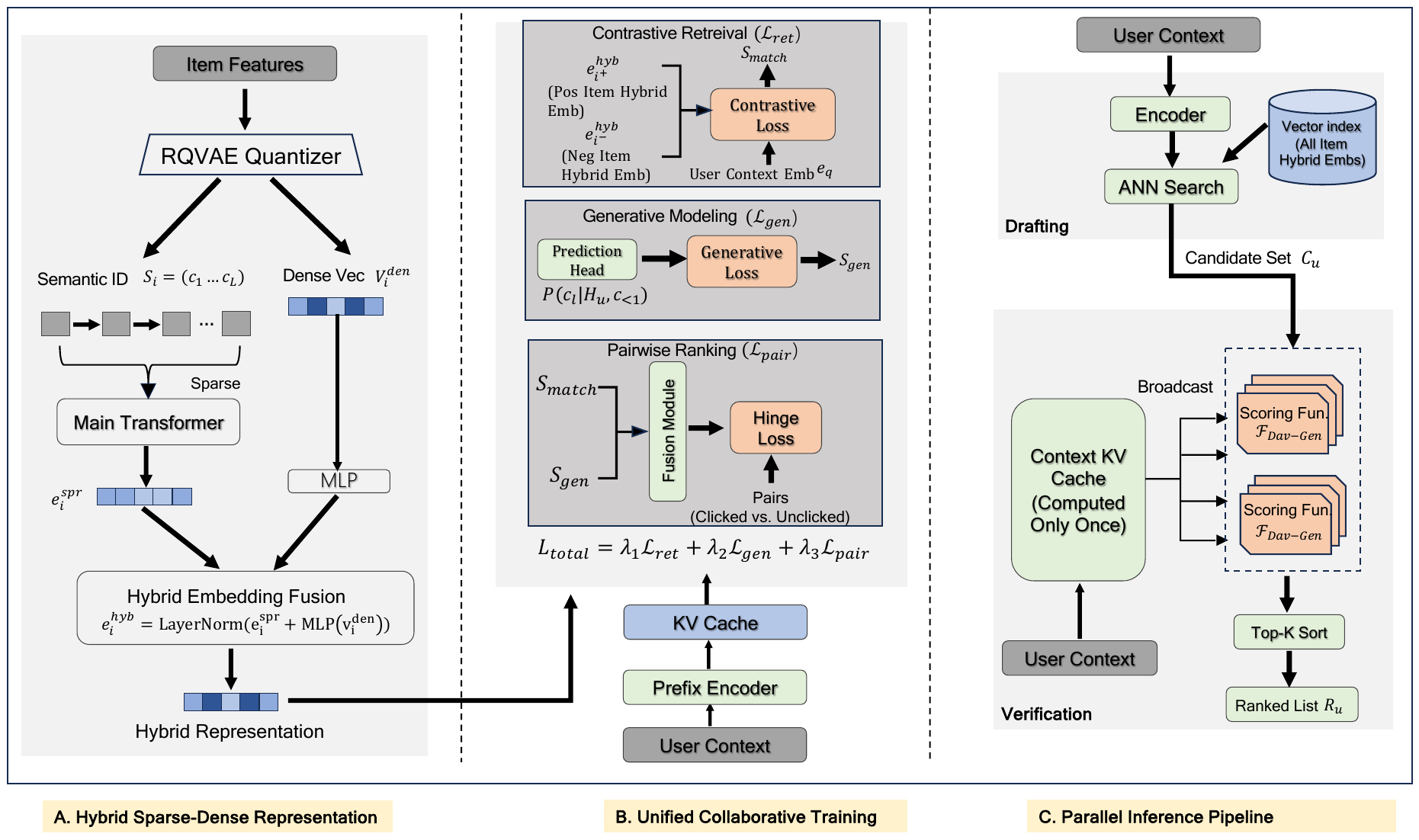}
    \caption{Overview of \textbf{DaV-Gen}.
    (a) \textbf{Hybrid Sparse-Dense Representation}: We construct a robust item embedding by fusing the fine-grained dense vector from the RQ-VAE encoder with the contextualized sparse Semantic ID.
    (b) \textbf{Unified Training Framework}: The model is jointly optimized with a composite loss (Contrastive + Generative + Pairwise) to align drafting and verification objectives within a single model.
    (c) \textbf{Parallel Inference Pipeline}: Adopting a ``Draft-and-Verify'' paradigm, DaV-Gen first drafts candidates via ANN search and then performs parallel verification in a single forward pass, significantly reducing latency compared to token-by-token generation.}
    \label{fig:framework}
\end{figure*}
Instead of relying on slow token-by-token generation, DaV-Gen reformulates the task into two synergistic operations within a single, end-to-end differentiable architecture (Figure~\ref{fig:framework}).
We systematically tackle the inherent trade-offs in current systems by addressing three fundamental design questions:

\textbf{1) How to balance representational efficiency and expressiveness?} (Figure~\ref{fig:framework}a)
Standard sparse tokens facilitate generation but often lose fine-grained semantics, while dense vectors excel at retrieval but lack structural hierarchy.
Our design resolves this by synthesizing a \textit{Hybrid Sparse-Dense Representation}.
By fusing the raw continuous signals from the RQ-VAE encoder with the contextualized discrete Semantic ID, we create a robust embedding that supports high-recall vector drafting without sacrificing the granular precision needed for downstream verification.

\textbf{2) How to align disparate objectives in a single model?} (Figure~\ref{fig:framework}b)
Solving the objective inconsistency of MCAs requires more than just sharing parameters; it demands a unified optimization landscape.
We construct a \textit{Collaborative Training Strategy} where the model is co-optimized for both contrastive discrimination and generative likelihood.
Crucially, a fusion-based pairwise objective acts as a bridge, ensuring that the drafting module's semantic space is perfectly aligned with the verification module's ranking preferences, forcing the two stages to reinforce rather than contradict each other.

\textbf{3) How to break the latency bottleneck of generative models?} (Figure~\ref{fig:framework}c)
The primary barrier to deploying GenIR is the serial dependency of auto-regressive decoding.
We dismantle this barrier by introducing a \textit{Parallel Inference Pipeline}.
By decoupling candidate acquisition from scoring, DaV-Gen first drafts candidates via efficient ANN search and then verifies the entire batch in a single forward pass.
This distinct architectural choice replaces the $O(L)$ complexity of sequential decoding with $O(1)$ parallel scoring, enabling real-time responsiveness.

The main contributions of this paper are summarized as follows:
\begin{itemize}
    \item \textbf{Draft-and-Verify Paradigm:} We propose DaV-Gen, a unified framework that refactors the generative retrieval paradigm from sequential token generation to parallel verification.
    This architecture fundamentally resolves the latency bottleneck of auto-regressive models, achieving industrial-grade inference efficiency that significantly outperforms traditional cascaded systems, while enabling precise list-length control.
    \item \textbf{Hybrid Sparse-Dense Representation:} We introduce a novel item representation that fuses fine-grained dense vectors from the RQ-VAE encoder with contextualized sparse Semantic IDs.
    This hybrid approach bridges the gap between efficient retrieval and precise semantic understanding, enhancing both the recall of the drafting stage and the accuracy of the verification stage.
    \item \textbf{Unified Optimization Framework:} We design a comprehensive training pipeline driven by a composite loss function.
    This strategy effectively eliminates the objective inconsistency of traditional cascades by optimizing a single model to simultaneously master coarse-grained drafting and fine-grained verification within a unified vector space.
\end{itemize}

\section{Methodology}

In this section, we first formalize the problem of generative retrieval and describe the conventional auto-regressive approach.
We then present our proposed DaV-Gen framework, detailing its comprehensive training process and efficient inference architecture.
\subsection{Formalism for Generative Retrieval}
Let $\mathcal{U}$ denote the set of users and $\mathcal{I}$ be the set of all items in the corpus.
Both search and recommendation share the same goal: retrieving relevant items based on a user's context.
We define the comprehensive user context as $\mathcal{C}_{ctx}$.
In Recommendation, the context is implicit: $\mathcal{C}_{ctx} = H_u = (\mathcal{S}_u, \mathcal{P}_u)$, where $\mathcal{S}_u = (i_1, \ldots, i_t)$ is the interaction history and $\mathcal{P}_u$ is the user profile.
In Search, the context is explicit: $\mathcal{C}_{ctx} = \{q, H_u\}$, driven primarily by the query $q$ and augmented by history $H_u$.
For notation simplicity and consistency, we use $H_u$ to denote this generalized context throughout the paper.
The foundation of modern generative recommendation is the concept of a \textbf{Semantic ID}.
Instead of using a single atomic identifier for each item, we first represent each item $i \in \mathcal{I}$ by its rich content features.
A quantizer model $\mathcal{Q}$, such as the pre-trained RQ-VAE~\cite{rajput2023recommender}, maps these features to a structured Semantic ID, which is a sequence of discrete tokens (or codewords) $S_i = (c_1, c_2, \ldots, c_L)$, where each token $c_j$ belongs to a finite codebook $\mathcal{C}$.
The task is thus to model the conditional probability $P(S_{i_{t+1}} | H_u)$, predicting the Semantic ID of the next relevant item $i_{t+1}$ given the context.
\subsection{Auto-Regressive Recommendation}
The mainstream approach to solving this task, as seen in models like TIGER, is purely auto-regressive.
Crucially, these models model the joint probability of a user's entire interaction sequence $\mathcal{S}_u = (i_1, \dots, i_T)$ as a strict causal chain.
The training objective minimizes the negative log-likelihood accumulated over all time steps in the sequence:
\begin{equation}
\mathcal{L}_{\text{AR}} = - \sum_{t=1}^{T} \sum_{l=1}^{L} \log P(c_l^{(t)} | i_{1}, \dots, i_{t-1}, c_{<l}^{(t)})
\label{eq:autoregressive}
\end{equation}
where $c_l^{(t)}$ denotes the $l$-th token of the item $i_t$ at step $t$, and the condition $i_{1}, \dots, i_{t-1}$ explicitly enforces the dependency on the entire preceding history.
This formulation necessitates a sequential, item-by-item decoding process, as the prediction of step $t$ strictly awaits the completion of step $t-1$, serving as the primary source of high inference latency.

\subsection{The DaV-Gen Training Process}
Our training process is a comprehensive pipeline designed to create an end-to-end model.
\subsubsection{Progressive Knowledge Injection}

The primary challenge in adopting a pre-trained language model (PLM) for generative retrieval lies in the semantic gap between the PLM's open-world knowledge and the domain-specific item representations (i.e., Semantic IDs).
Since the discrete tokens constituting a Semantic ID are randomly initialized, directly optimizing the entire model can lead to catastrophic forgetting of the PLM's linguistic capabilities.
To address this, we design a robust strategy that progressively injects item-specific knowledge into the backbone model via a self-supervised reconstruction task.
In this task, the model is trained to bidirectionally predict an item's Semantic ID given its textual attributes and vice versa.
\begin{enumerate}
    \item \textbf{Embedding Alignment.} 
    In the initial stage, we focus solely on aligning the newly introduced Semantic ID tokens with the pre-existing semantic space of the PLM.
    We freeze all parameters of the transformer backbone, making only the embedding matrix of new tokens trainable.
    By projecting the textual descriptions of items into the model's latent space and optimizing the embeddings of their corresponding Semantic IDs to match these representations, we establish a coarse-grained alignment.
    This warm-up process ensures that the Semantic ID tokens acquire meaningful vector representations without destabilizing the carefully pre-trained weights of the base model.
    \item \textbf{Deep Semantic Integration.} 
    Once the token embeddings are preliminarily aligned and stabilized, we proceed to the second stage by unfreezing all model parameters.
    The pre-training task continues in a full-parameter fine-tuning regime.
    This phase allows the model to learn complex, non-linear interactions between the item's hierarchical structure (captured by the Semantic ID sequence) and its semantic content.
    It enables the backbone to deeply integrate the specific item knowledge with its general world knowledge, effectively transforming the PLM from a general-purpose text generator into a domain-expert recommender system.
\end{enumerate}

\subsubsection{Hybrid Sparse-Dense Item Representation}
We draw inspiration from the COBRA framework~\cite{yang2025sparse} to leverage the mutual reinforcement of sparse and dense signals.
However, distinct from COBRA, which models them as an alternating sequence for step-by-step auto-regressive generation, we fuse them into a single unified vector.
This design choice is critical for our ``Draft-and-Verify'' paradigm: it compresses the item's multimodal semantics into a fixed-length embedding compatible with Maximum Inner Product Search (MIPS), enabling the high-speed drafting that COBRA's sequential architecture cannot directly support.

Specifically, the hybrid embedding $e_i^{\text{hyb}}$ has two parts:
\begin{itemize}
    \item \textbf{Dense Content Vector ($v_i^{\text{den}}$):} Derived directly from the RQ-VAE's encoder output before quantization. It preserves the fine-grained, non-discrete semantic features (e.g., specific textual descriptions) often lost in tokenization.
    \item \textbf{Sparse Structural Vector ($e_i^{\text{spr}}$):} To capture the hierarchical category information of the Semantic ID $S_i$, we feed the token sequence into the backbone and compute $e_i^{\text{spr}}$ via mean-pooling the final hidden states.
\end{itemize}

These components are fused via a projection layer to form the final item anchor:
\begin{equation}
e_i^{\text{hyb}} = \text{LayerNorm}(e_i^{\text{spr}} + \text{MLP}(v_i^{\text{den}}))
\label{eq:hybrid_rep}
\end{equation}
where $\text{MLP}$ aligns the dimensions. This representation serves as the indexable vector for drafting and the semantic anchor for verification.

\subsubsection{Supervised Fine-Tuning (SFT) Optimization Strategy}
During SFT, the model is optimized with a composite loss function.
The total loss $\mathcal{L}_{\text{total}}$ is a weighted sum of three key components:
\begin{equation}
\mathcal{L}_{\text{total}} = \lambda_1\mathcal{L}_{\text{ret}} + \lambda_2\mathcal{L}_{\text{gen}} + \lambda_3\mathcal{L}_{\text{pair}}
\label{eq:total_loss}
\end{equation}
where $\lambda_1$, $\lambda_2$, and $\lambda_3$ are hyper-parameters.
This composite objective equips the model with three essential skills: $\mathcal{L}_{\text{ret}}$ teaches the model to produce effective embeddings for drafting;
$\mathcal{L}_{\text{gen}}$ imparts a deep contextual understanding of item sequences; and $\mathcal{L}_{\text{pair}}$ directly optimizes the final verification capability.
We detail each component below.

\paragraph{(1) Contrastive Drafting Loss ($\mathcal{L}_{\text{ret}}$).}
This loss is the cornerstone of our efficient drafting stage, designed to structure the embedding space for fast and accurate candidate generation.
We calculate this loss over the training batch $\mathcal{B}$. For each user context $H_u$ in the batch, we define the candidate set $\mathcal{C}_u = \{i^+\} \cup \{i^-_j\}$ as the union of the positive item $i^+$ and the set of in-batch negative items.
The contrastive loss is formulated as the average over the batch:
\begin{equation}
\mathcal{L}_{\text{ret}} = - \frac{1}{|\mathcal{B}|} \sum_{u \in \mathcal{B}} \log \frac{\exp(\text{sim}(e_q, e_{i^+}^{\text{hyb}}) / \tau)}{\sum_{k \in \mathcal{C}_u} \exp(\text{sim}(e_q, e_{k}^{\text{hyb}}) / \tau)} 
\label{eq:ret_loss}
\end{equation}
where $e_q$ is the user embedding derived from $H_u$, $e_{k}^{\text{hyb}}$ denotes the hybrid embedding for item $k$, $\text{sim}(\cdot, \cdot)$ is the cosine similarity, and $\tau$ is a temperature hyper-parameter that controls the sharpness of the distribution.

This loss explicitly teaches the model the concept of semantic relevance.
By rewarding the proximity of a user representation to its positive item while penalizing its proximity to numerous negative items, we compel the model to organize the high-dimensional embedding space into meaningful, semantically coherent clusters.
The superiority of this approach is twofold. Firstly, it produces embeddings that are directly optimized for MIPS.
This is the critical property that allows us to use highly efficient Approximate Nearest Neighbor (ANN) search libraries (e.g., FAISS) for the initial drafting stage, which directly solves the high-latency problem of purely generative models.
Secondly, by learning from a diverse set of negatives, the model gains a robust understanding of item-to-item relationships, enhancing its ability to generalize and recommend novel yet relevant items, thus improving the quality of the candidate set passed to the subsequent verification phase.
\paragraph{(2) Generative Loss ($\mathcal{L}_{\text{gen}}$).}
This loss leverages a generative task to capture a deep, contextual understanding of items.
Unlike standard auto-regressive recommenders that model dependencies between items, our approach treats each candidate item independently.
However, we explicitly retain the auto-regressive dependency within the tokens of each item's Semantic ID ($c_1, \ldots, c_L$) to capture its internal hierarchical structure.
We formulate the objective over the entire training batch $\mathcal{B}$, where each sample consists of a user context $H_u$ and the corresponding ground-truth positive item $i$.
The loss is defined as the average negative log-likelihood across the batch:
\begin{equation}
\mathcal{L}_{\text{gen}} = - \frac{1}{|\mathcal{B}|} \sum_{(H_u, i) \in \mathcal{B}} \sum_{l=1}^{L} \log P(c_l | H_u, c_{<l})
\label{eq:gen_loss}
\end{equation}
where $P(c_l | H_u, c_{<l})$ is the predicted probability for the $l$-th token of item $i$, conditioned on the user history $H_u$ and the preceding tokens $c_{<l}$ of the same item.

This design extracts a verification signal that is orthogonal to simple vector similarity.
While the contrastive loss answers ``which items are semantically similar?'', the generative loss answers ``how plausible is this specific item in this user's context?''.
It forces the model to move beyond geometric proximity and learn the fine-grained compositional semantics of an item (as encoded by its RQ-VAE tokens) and how they align with a user's dynamic preferences.
The resulting negative log-likelihood serves as a powerful generative score that enables a sophisticated and accurate final verification decision when fused with the matching score.

To efficiently compute this loss across a large batch of candidates, we implement a Broadcasted Prefix Caching mechanism.
A naive implementation would necessitate concatenating the long context $H_u$ with every candidate item, leading to redundant computations where the context is re-encoded $N$ times.
Instead, we compute the Key-Value (KV) states of $H_u$ once per batch and broadcast them to serve as a shared memory prefix for all candidates.
The model then computes the likelihood of item tokens by attending to this pre-computed cache, effectively reducing the context encoding complexity from $O(N \cdot |H_u|)$ to $O(|H_u|)$.

\paragraph{(3) Pairwise Ranking Loss ($\mathcal{L}_{\text{pair}}$).}
This loss directly optimizes the model's final verification capability using a fused score and explicit user feedback.
First, we define the final score $\sigma(i)$ for a candidate item $i$ as the output of a fusion network $f_{\text{fusion}}$, which takes two inputs: a matching score $s_{\text{match}}$ and a generative score $s_{\text{gen}}$.
\begin{equation}
\sigma(i) = f_{\text{fusion}}(s_{\text{match}}(i), s_{\text{gen}}(i))
\label{eq:fusion}
\end{equation}
where $s_{\text{match}}(i) = \text{sim}(e_q, e_i^{\text{hyb}})$ and $s_{\text{gen}}(i)$ is derived from the item's generative loss.
The fusion network allows the model to learn the optimal, non-linear combination of these two signals.
The model is then trained on this score using a pairwise hinge loss, averaged over valid pairs constructed from the batch.
We construct two distinct sets of pairs:
\begin{enumerate}
    \item Hard pairs $\mathcal{S}_{\text{hard}}$: Pairs of clicked items $i$ and unclicked items $j$ from the same session.
    \item Calibration pairs $\mathcal{S}_{\text{rand}}$: Pairs of exposed items $i$ and randomly sampled negative items $k$.
\end{enumerate}
The objective is to ensure the score of a preferred item is higher than that of a non-preferred item by at least a margin $m$:
\begin{equation}
\begin{aligned}
\mathcal{L}_{\text{pair}} = & \quad \frac{1}{|\mathcal{S}_{\text{hard}}|} \sum_{(i,j) \in \mathcal{S}_{\text{hard}}} [\max(0, \sigma(j) - \sigma(i) + m)] \\
& + \frac{1}{|\mathcal{S}_{\text{rand}}|} \sum_{(i,k) \in \mathcal{S}_{\text{rand}}} [\max(0, \sigma(k) - \sigma(i) + m)]
\end{aligned}
\label{eq:pair_loss}
\end{equation}
The motivation for using these two types of pairs is to create a robust and well-calibrated verifier.
The first set of pairs (clicked vs. unclicked) presents the model with hard negatives.
This teaches the model to make fine-grained distinctions among items that are all relevant enough to be shown to the user, directly optimizing for engagement.
The second set of pairs (exposed vs. random) provides a crucial global calibration.
It teaches the model that any item deemed relevant enough for exposure should be scored significantly higher than a random item from the corpus.
This ensures a sensible scoring baseline and improves the overall robustness of the verification function.
\subsection{Parallelized Generative Inference}
Our comprehensive training process enables a highly efficient two-stage inference architecture that is well-suited for online serving.
\subsubsection{Stage 1: Candidate Drafting}
The inference process begins with a fast candidate drafting step.
Given a user's context $H_u$, the framework first computes the corresponding query/context embedding $e_q$. Let $\mathcal{E}_{\mathcal{I}} = \{e_i^{\text{hyb}} |
i \in \mathcal{I}\}$ be the pre-computed set of hybrid embeddings for all items in the corpus, stored in an efficient vector index.
The drafting stage produces a candidate set $\mathcal{C}_u$ of size $N$ using an Approximate Nearest Neighbor (ANN) search:
\begin{equation}
\mathcal{C}_u = \text{ANN}(e_q, \mathcal{E}_{\mathcal{I}}, N)
\label{eq:ann}
\end{equation}
where $\text{ANN}(\cdot)$ represents the fast vector search operation.
This step is non-generative and returns a fixed-size candidate list almost instantaneously.
\subsubsection{Stage 2: Parallel Verification}
The $N$ retrieved candidates from $\mathcal{C}_u$ are then passed to our trained DaV-Gen model for scoring.
Crucially, to ensure low latency, we leverage the Broadcasted Prefix Caching mechanism introduced in the training phase.
Instead of concatenating the user context with each candidate repeatedly, we compute the context's KV cache once and broadcast it to initialize the attention states for all $N$ candidates.
The model then processes the tokens of all $N$ candidates in a single parallel forward pass, conditioned on this cached prefix.
Let $\mathcal{F}_{\text{DaV-Gen}}$ be the learned scoring function. For every candidate item $i \in \mathcal{C}_u$, the score is computed as:
\begin{equation}
\text{score}(i) = \mathcal{F}_{\text{DaV-Gen}}(i | \text{Cache}(H_u)) \quad \forall i \in \mathcal{C}_u
\label{eq:scoring}
\end{equation}
The candidates are then sorted in descending order based on this score to form a ranked list $\mathcal{L}_u$.
This design decouples the context length from the candidate set size, ensuring that the inference latency grows only marginally with the number of verified items.
The final recommendation list, $\mathcal{R}_u$, consists of the top-K items from this list:
\begin{equation}
\mathcal{R}_u = \text{Top-K}(\mathcal{L}_u)
\label{eq:topk}
\end{equation}

This architecture directly solves the challenges of prior approaches.
It sidesteps the high-latency auto-regressive bottleneck by replacing sequential generation with fast drafting and parallel verification.
Furthermore, since the number of items to draft ($N$) and display ($K$) are fixed parameters, it offers precise, deterministic control over the resulting list length.
Finally, because a single, unified model is trained for both drafting and verification, the objective inconsistency of classic cascaded systems is eliminated.
\section{Experiments}

To validate the effectiveness and generality of DaV-Gen, we conduct extensive experiments covering both recommendation and search scenarios.
Our evaluation aims to answer the following research questions:
\textbf{RQ1 (Model Accuracy):} Does DaV-Gen outperform state-of-the-art models on widely used benchmarks?
\textbf{RQ2 (Industrial Scalability):} Can DaV-Gen handle large-scale industrial scenarios effectively compared to traditional methods?
\textbf{RQ3 (Online Business Value):} Does the framework deliver real-world business improvements in an online production environment?
\textbf{RQ4 (Ablation):} What is the contribution of our proposed innovative components in  DaV-Gen?

\subsection{Experimental Setup}

\subsubsection{Datasets}
Following recent works~\cite{wang2024generative,xing2025reg4rec}, we perform experiments on three widely used recommendation benchmarks: \textbf{Amazon Beauty}, \textbf{Amazon Sports}, and \textbf{Yelp}.
To verify the model's scalability in Search, we include a large-scale industrial dataset, \textbf{Ind-Search}, collected from a leading video search engine, containing over 100M items and 500M search logs.
For all datasets, we adopt the leave-one-out strategy for evaluation.
\subsubsection{Baselines}
We compare DaV-Gen with representative methods:
\begin{itemize}
    \item \textbf{Discriminative Models:} \textbf{SASRec}~\cite{ssr2018kang}, \textbf{BERT4Rec}~\cite{sun2019bert4rec}, \textbf{HGN}~\cite{ma2019hierarchical}, and the industrial strong baseline \textbf{MoE} (a Mixture-of-Experts based DNN model for CTR prediction) serving online.
    \item \textbf{Generative \& Retrieval Models:} \textbf{TIGER}~\cite{rajput2023recommender}, \textbf{LC-Rec}~\cite{wang2024generative}, \textbf{OneRec}~\cite{deng2025onerec}, and the state-of-the-art dense retriever \textbf{gte-qwen-7b} (fine-tuned).
\end{itemize}

\subsubsection{Implementation Details}
For fair comparison, the embedding dimension is set to 64 for all models.
For generative models (TIGER, OneRec, DaV-Gen), we use a codebook of 5 layers of hierarchical quantization.
DaV-Gen is trained with the Adam optimizer with a learning rate of $1e-5$ and a batch size of 256. 
\subsection{Performance Comparison (RQ1)}

Table~\ref{tab:main_results} reports the performance on public recommendation datasets. We observe the following:

\begin{table*}[t]
\centering
\resizebox{\textwidth}{!}{
\begin{tabular}{l|cc|cc|cc}
\toprule
\multirow{2}{*}{\textbf{Method}} & \multicolumn{2}{c|}{\textbf{Amazon Beauty}} & \multicolumn{2}{c|}{\textbf{Amazon Sports}} & \multicolumn{2}{c}{\textbf{Yelp}} \\
 & Recall@10 & NDCG@10 & Recall@10 & NDCG@10 & Recall@10 & NDCG@10 \\
\midrule
SASRec & 0.0624 & 0.0385 & 0.0482 & 0.0295 & 0.0681 & 0.0412 \\
BERT4Rec & 0.0638 & 0.0392 & 0.0495 & 0.0301 & 0.0695 & 0.0425 \\
HGN & 0.0705 & 0.0441 & 0.0571 & 0.0358 & 0.0738 & 0.0462 \\
\midrule
TIGER & 0.0698 & 0.0435 & 0.0552 & 0.0348 & 0.0725 & 0.0455 \\
LC-Rec & 0.0715 & 0.0452 & 0.0568 & 0.0355 & 0.0742 & 0.0468 \\
OneRec & \underline{0.0732} & \underline{0.0465} 
& \underline{0.0588} & \underline{0.0370} & \underline{0.0765} & \underline{0.0481} \\
\midrule
\textbf{DaV-Gen (Ours)} & \textbf{0.0752} & \textbf{0.0481} & \textbf{0.0605} & \textbf{0.0382} & \textbf{0.0784} & \textbf{0.0495} \\
\textit{Improv.} & +2.73\% & +3.44\% & +2.89\% & +3.24\% & +2.48\% & +2.91\% \\
\bottomrule
\end{tabular}
}
\caption{Performance comparison on three recommendation datasets.
The best results are highlighted in \textbf{bold}, and the second best are \underline{underlined}.
Improvement is calculated relative to the strongest baseline.}
\label{tab:main_results}
\end{table*}

\textbf{1.
Superiority over Discriminative Models:} Generative methods generally outperform SASRec and BERT4Rec.
This demonstrates that directly modeling the generation probability of semantic IDs captures collaborative signals more effectively.
\textbf{2.
Comparison with Unified Frameworks:} OneRec performs significantly better than TIGER and RPG, validating the effectiveness of end-to-end training models.
However, DaV-Gen further surpasses OneRec. While OneRec relies on auto-regressive decoding, DaV-Gen utilizes a ``Draft-and-Verify'' mechanism that globally scores candidates, leading to more robust verification performance.
\subsection{Evaluation on Industrial Search (RQ2)}

To assess the industrial scalability of DaV-Gen, we evaluated it on the Ind-Search dataset against two strong production baselines: the fine-tuned dense retriever gte-qwen-7b and the online ranking model MoE.
We report Recall@50, NDCG@10, and MRR@10.

\begin{figure}[t]
    \centering
    \includegraphics[width=1\columnwidth]{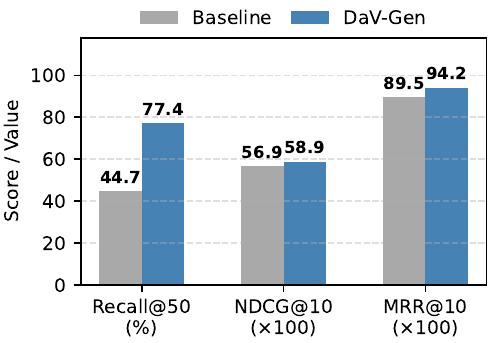}
    \caption{Performance comparison on Ind-Search.
    Metrics for ranking (NDCG, MRR) are scaled by $\times 100$ for visualization.}
    \label{fig:search_metrics_compact}
\end{figure}

Figure~\ref{fig:search_metrics_compact} summarizes the comparison.
\textbf{Retrieval (Recall@50):} DaV-Gen achieves a dominant 77.4\%, outperforming the dense retriever baseline (44.7\%) by over 30 absolute percentage points.
This confirms that our Hybrid Sparse-Dense Representation effectively captures the complex semantics in video search queries that pure dense vectors miss.
\textbf{Ranking (NDCG \& MRR):} DaV-Gen consistently surpasses the discriminative MoE baseline.
It improves NDCG@10 from 0.569 to 0.589 and boosts MRR@10 from 0.895 to 0.942.
The high MRR score specifically highlights DaV-Gen's ability to place the best result at top positions, which is critical for optimizing user search experience.
\subsection{Online A/B Testing (RQ3)}

We deployed DaV-Gen in the production environment of a leading video provider to serve live search traffic.
The experiment involved  millions of users and was conducted over one week.
We focus on three core business metrics reflecting user engagement, conversion efficiency and search satisfaction:

\begin{itemize}
    \item \textbf{Avg. Time Spent per User (ATS):} The total time spent divided by the number of exposed users ($\text{ATS} = \frac{\sum \text{Time Spent}}{\text{Exposed UV}}$).
    This is the primary metric for user stickiness.
    \item \textbf{User Conversion Rate (UCVR):} The ratio of users who performed at least one video play to the total exposed users ($\text{UCVR} = \frac{\text{Play UV}}{\text{Exposed UV}}$), reflecting the feed's ability to convert exposure into consumption.
    \item \textbf{Avg. Successful Searches (ASS):} The number of search queries with at least one click divided by exposed users ($\text{ASS} = \frac{\text{Hit Search Volume}}{\text{Exposed UV}}$), indicating the efficiency of satisfying user intent within a session.
\end{itemize}

\begin{table}[t]
\centering
\resizebox{0.95\columnwidth}{!}{
\begin{tabular}{l|cc}
\toprule
\textbf{Metric} & \textbf{Improvement} & \textbf{Significance} \\
\midrule
\textbf{Avg. Time Spent per User} & +2.09\% & \textbf{$p < 0.05$} \\
\textbf{User Conversion Rate} & +0.47\% & $p < 0.05$ \\
\textbf{Avg. Successful Searches} & +0.31\% & $p < 0.05$ \\
\bottomrule
\end{tabular}
}
\caption{Online A/B Testing Results}
\label{tab:online_ab}
\end{table}

Table~\ref{tab:online_ab} reports the experimental results relative to the highly optimized production baseline.
DaV-Gen yields a +2.09\% uplift in ATS, suggesting that our fine-grained verification mechanism significantly enhances user stickiness by ensuring the most engaging content is prioritized.
Furthermore, the consistent gains in UCVR (+0.47\%) and ASS (+0.31\%) demonstrate the model's superior capability in capturing user intent and facilitating content discovery.
\paragraph{Latency Analysis.} 
Efficiency is critical for online serving. We explicitly compare the inference latency of DaV-Gen against two representative paradigms:
1) \textbf{Pure Generative Models:} Standard auto-regressive models (e.g., TIGER) suffer from prohibitive latency ($\approx \mathbf{3s}$) due to the sequential item-by-item decoding process, making them impractical for real-time search.
2) \textbf{Traditional Cascade Structure:} The current production baseline employs a complex funnel (Query Processing $\to$ Recall $\to$ Pre-Ranking $\to$ Ranking), resulting in an accumulated latency of $\approx \mathbf{130ms}$.
3) \textbf{DaV-Gen (Ours):} By adopting a unified ``Draft-and-Verify'' strategy and leveraging Parallel Scoring, DaV-Gen reduces the latency to $\approx \mathbf{70ms}$.
This represents a 2.5$\times$ speedup compared to the traditional cascade and is orders of magnitude faster than pure generative approaches, demonstrating superior suitability for high-concurrency industrial environments.

\subsection{Ablation Study (RQ4)}
To verify the robustness of our design, we extend the ablation study to two datasets: Amazon Beauty and Amazon Sports.
We investigate the contribution of key components by evaluating the following variants:
\begin{itemize}
    \item \textbf{w/o Hybrid Rep.:} Removes the dense semantic vectors, relying solely on sparse ID matching.
    \item \textbf{w/o Prog. Inject.:} Skips  Embedding Alignment of the progressive knowledge injection, directly proceeding to the full-parameter Deep Semantic Integration.
    \item \textbf{w/o Gen. Loss ($\mathcal{L}_{\text{gen}}$):} Removes the token-level generative objective, training the model only with contrastive and pairwise ranking losses.
    \item \textbf{w/o Fusion Loss ($\mathcal{L}_{\text{pair}}$):} Removes the fusion-based pairwise ranking objective, relying solely on the sum of contrastive and generative scores for inference.
\end{itemize}

\begin{table}[t]
\centering
\resizebox{\columnwidth}{!}{
\begin{tabular}{l|cc|cc}
\toprule
\multirow{2}{*}{\textbf{Variant}} & \multicolumn{2}{c|}{\textbf{Amazon Beauty}} & \multicolumn{2}{c}{\textbf{Amazon Sports}} \\
 & Recall@10 & NDCG@10 & Recall@10 & NDCG@10 \\
\midrule
\textbf{DaV-Gen (Full)} & \textbf{0.0752} & \textbf{0.0481} & \textbf{0.0605} & \textbf{0.0382} \\
\midrule
w/o Hybrid Rep. & 0.0710 & 0.0445 & 0.0569 & 0.0355 \\
w/o Prog. Inject. & 0.0722 & 0.0458 & 0.0580 & 0.0366 \\
w/o Gen. Loss & 0.0718 & 0.0454 & 0.0575 & 0.0362 \\
w/o Fusion Loss & 0.0728 & 0.0462 & 0.0588 & 0.0371 \\
\bottomrule
\end{tabular}
}
\caption{Ablation study on Amazon Beauty and Sports datasets.}
\label{tab:ablation}
\end{table}

The results across both datasets, summarized in Table~\ref{tab:ablation}, provide consistent and critical insights:

(1) Removing the dense vectors leads to the most significant performance drop (e.g., -5.6\% Recall on Beauty). 
This validates that sparse semantic tokens alone are insufficient; combining them with dense embeddings significantly enriches the model's expressiveness and drafting coverage.
(2) The variant skipping the alignment stage (\textit{w/o Prog. Inject.}) suffers a notable decline. 
This demonstrates that the ``semantic gap'' between the PLM and Semantic IDs requires a progressive bridge. 
Without the initial embedding alignment to map ID tokens into the PLM's latent space, the subsequent full-parameter learning struggles to converge optimally.
(3) The removal of $\mathcal{L}_{\text{gen}}$ results in clear degradation. 
This proves that the token-level generative objective is not merely auxiliary; it forces the model to learn fine-grained sequential dependencies and internal item semantics that contrastive learning alone cannot capture.
(4) The collaborative matching fusion loss ensures that the generation objective aligns with the verification metric. 
Without it, we observe a consistent drop in NDCG, confirming its role in refining the final verification order.

\section{Related Works}
Existing approaches in recommender systems generally fall into two paradigms: the established MCAs and the emerging GenIR.
DaV-Gen bridges these worlds, integrating the efficiency of cascades with the unified expressiveness of generative models.

\textbf{Discriminative Multi-stage Cascading Architectures.}
MCAs have long been the industrial standard, handling massive corpora by decomposing the ranking problem into sequential stages~\cite{qin2022rankflow}.
The retrieval stage prioritizes high-recall candidate selection via Maximum Inner Product Search (MIPS).
This evolved from early Collaborative Filtering to deep dual-encoder models like DSSM~\cite{dssm2013huang} and YoutubeDNN~\cite{covington2016deep}, and later to sequential models like SASRec~\cite{ssr2018kang} which capture dynamic user preferences.
The subsequent ranking stage employs complex interaction-focused architectures, such as DIN~\cite{zhou2018deep} and SIM~\cite{sim2020pi}, to capture fine-grained interests.
However, the strict separation between these stages creates an ``objective inconsistency'' problem: the retriever's training objective is often misaligned with the ranker's, leading to irreversible error propagation~\cite{hron2021component}.

\textbf{Generative Retrieval and Ranking.}
GenIR reformulates IR as a unified sequence-to-sequence generation task, directly predicting target item identifiers.
Early approaches like P5~\cite{geng2022recommendation} and M6-Rec~\cite{cui2022m6} leveraged Pre-trained Language Models (PLMs) to process raw text tokens.
While semantically rich, these methods incurred prohibitive computational costs and struggled with context limits.
To address this, the field shifted towards discrete Semantic IDs derived from Residual Quantized VAEs (RQ-VAE).
Models like TIGER~\cite{rajput2023recommender} and LC-Rec~\cite{wang2024generative} generate these hierarchical tokens auto-regressively, significantly improving the modeling of collaborative signals.
Despite their effectiveness, the sequential dependency of token-by-token decoding imposes a severe latency bottleneck~\cite{rajput2023recommender,wang2024generative}.
Recent research has thus pivoted to address these limitations from different angles: RPG~\cite{hou2025rpg} proposes parallel generation to break the decoding bottleneck; TBGRecall~\cite{liang2025tbgrecall} and REG4Rec~\cite{xing2025reg4rec} enhance the model's ability to handle complex sessions and reasoning paths; and OneRec~\cite{deng2025onerec} focuses on structurally unifying retrieval and ranking to solve objective inconsistency.
Distinct from these efforts, DaV-Gen proposes a  ``Draft-and-Verify'' paradigm, which uniquely combines the inference speed of retrieval-based drafting with the precision of parallel generative verification.
\section{Conclusion}

We presented DaV-Gen, refactoring generative retrieval from sequential generation to parallel verification.
Our Hybrid Sparse-Dense Representation and ``Draft-and-Verify'' mechanism successfully resolve objective inconsistency in cascades and decoding latency in generative models.
Extensive experiments confirm its superior accuracy and efficiency.
Future work will extend the modality-agnostic Dense Content Vector with visual or audio embeddings for richer multimodal fusion.

\section*{Contribution Statement}
Meng Zhao and Chunmei Liu contributed equally to this work and are co-first authors. Qinyong Wang is the corresponding author.

\bibliographystyle{named}
\bibliography{ijcai26}

\end{document}